\documentclass[%
 preprint, amsmath,amssymb,
 aps, physrev,superscriptaddress]{revtex4-2}
\usepackage{braket}
\usepackage{graphicx}
\usepackage{cancel}
\usepackage{xcolor}
\usepackage{dcolumn}
\usepackage{bm}
\usepackage[colorlinks=true, allcolors=blue]{hyperref}
\usepackage[normalem]{ulem}
\graphicspath{ {./Figures/} }

\begin{document}


\title{\textbf{Gate-tunable spectrum and charge dispersion mitigation in a graphene superconducting qubit} 
}%

\author{Nicolas Aparicio}
\author{Simon Messelot}
\author{Edgar Bonet-Orozco}
\author{Eric Eyraud}
\affiliation{Univ. Grenoble Alpes, CNRS, Grenoble INP, Institut N\'eel, 38000 Grenoble, France}
\author{Kenji Watanabe}
\affiliation{Research Center for Electronic and Optical Materials, National Institute for Materials Science, 1-1 Namiki, Tsukuba 305-0044, Japan}
\author{Takashi Taniguchi}
\affiliation{Research Center for Materials Nanoarchitectonics, National Institute for Materials Science,  1-1 Namiki, Tsukuba 305-0044, Japan}
\author{Johann Coraux}
\affiliation{Univ. Grenoble Alpes, CNRS, Grenoble INP, Institut N\'eel, 38000 Grenoble, France}
\author{Julien Renard}
\affiliation{Univ. Grenoble Alpes, CNRS, Grenoble INP, Institut N\'eel, 38000 Grenoble, France}

\begin{abstract}

Controlling the energy spectrum of quantum-coherent superconducting circuits, i.e. the energies of excited states, the circuit anharmonicity and the states' charge dispersion, is essential for designing performant qubits. This control is usually achieved by adjusting the circuit's geometry. In-situ control is traditionally obtained via an external magnetic field, in the case of tunnel Josephson junctions. More recently, semiconductor-weak-links-based Josephson junctions have emerged as an alternative building block with the advantage of tunability via the electric-field effect. Gate-tunable Josephson junctions have been succesfully integrated in superconducting circuits using for instance semiconducting nanowires or two-dimensional electron gases. In this work we demonstrate, in a graphene superconducting circuit, a large gate-tunability of qubit properties: frequency, anharmonicity and charge dispersion. We rationalize these features using a model considering the transmission of Cooper pairs through Andreev bound states. Noticeably, we show that the high transmission of Cooper pairs in such weak link strongly suppresses the charge dispersion. Our work illustrates the potential for graphene-based qubits as versatile building-blocks in advanced quantum circuits.

\end{abstract}

\maketitle

\section{\label{sec:intro}Introduction\protect}

Microwave superconducting quantum circuits rely on the existence of a lossless nonlinear inductive element: the Josephson junction. This element is key to encode quantum information in a qubit, to perform parametric amplification at the quantum limit or to generate non-classical states of light. The most mature realizations of such circuits are arguably based on Josephson tunnel junctions \cite{Macklin,Somoroff2021} but alternatives, such as disordered superconductors, have been explored for instance to promote wave mixing in a travelling wave parametric amplifier \cite{Eom2012}, or to build transmon and fluxonium qubits \cite{PhysRevX.10.031032,Grunhaupt2019}.
Another alternative is to revisit the junction itself and replace the tunnel barrier by a conducting material. If its conduction can be tuned by the electric-field effect, this would allow to suppress the need for magnetic field tunability, which is traditionally used when dealing with tunnel Josephson junctions. 

Using such so-called hybrid junctions, gate-tunable qubits \cite{Larsen2015, De_lange, Casparis2018} and parametric amplifiers \cite{Guilliam,Sarkar, PhanD, Hao2024} have been demonstrated. Original demonstrations were made  with semiconducting nanowires as the weak link but other materials were also successfully employed since then, especially planar materials, which are amenable to a variety of patterning options. Accordingly, two-dimensional (2D) electron gases have been used, including semiconducting quantum wells \cite{Strickland2023,Kiyooka2025-uf} or the semi-metal graphene \cite{Jan2018}. In all these qubits, it was for instance possible to tune, with the gate voltage, the frequency of the transition to the first excited state by one to few GHz \cite{Casparis2016,Jan2018,Strickland2023}.

These hybrid junctions expectedly operate in a regime much different from that typical of Josephson tunnel junctions -- Cooper pairs are simply not transported the same way across them. This can have strong consequences on the qubit properties. In this work, we present the detailed spectroscopy of a charge qubit based on a gate-tunable graphene Josephson junction. We extract the energy spectrum, its anharmonicity, and the levels' charge dispersion. We analyze these characteristics using the Hamiltonian of a superconducting quantum point contact that takes into account the gate-dependent transmission of different channels as well as Andreev bound states from both branches across the superconducting gap. We accordingly relate the measured quantities to the offset charge as well as the gate-tunable Josephson energy and junction's transmission. We find that the large transmission reduces the charge dispersion, which could allow the design of highly-coherent qubits in the future.


\section{\label{sec:design}Circuit design, fabrication and methods\protect}

\begin{figure}[!htb]
\includegraphics{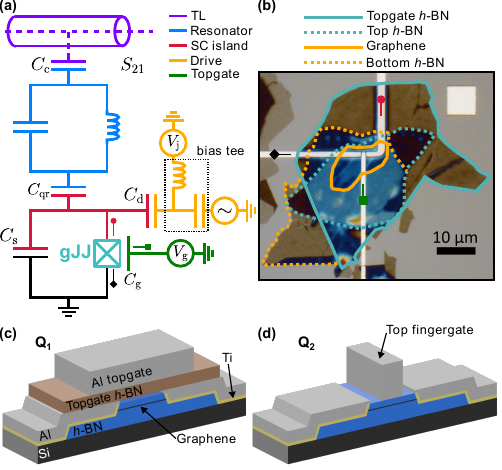}
\caption{\label{fig:device} Devices description. (a) Electrical circuit diagram of a charge qubit made of a graphene Josephson junction (gJJ) shunted to ground by a $C_\mathrm{S}$ capacitance, and coupled to a readout resonator by a $C_\mathrm{qr}$ capacitance, a drive line by a $C_\mathrm{d}$ capacitance, and a gate line by a $C_\mathrm{g}$ capacitance to control conduction through graphene. The qubit superconducting island is defined by the superconducting leads (red) between the junction and the $C_\mathrm{S}$ capacitance. The qubit is read out by probing the transmission scattering parameter, $S_\mathrm{21}$, of a transmission line (TL) coupled to a microwave resonator by a $C_\mathrm{c}$ capacitance. (b) Optical micrograph of the \textit{h}-BN/graphene/\textit{h}-BN heterostructure of device Q$_\mathrm{1}$. (c,d) Schematics of the top-gated gJJs side-contacted to the Ti/Al superconducting electrodes. In device Q$_\mathrm{1}$ (c), a third \textit{h}-BN layer is deposited on top of the electrodes as a gate dielectric to build an Al topgate. In device Q$_\mathrm{2}$ (d), the top \textit{h}-BN of the heterostructure is directly used as a gate dielectric and a narrow (100 nm-wide) Al electrode is deposited in between the superconducting electrodes.}
\end{figure}

Measurements were performed in a dilution refrigerator (base temperature $\sim$ 25 mK) on two different devices, each consisting of [see Fig. \ref{fig:device}(a)] a graphene-based Josephson junction shunted to ground by a large capacitance ($C_{\mathrm{S}}$), i.e. a charge-qubit-like circuit. This circuit is coupled to a coplanar $\lambda / 2$ waveguide (CPW) readout resonator via a $C_\mathrm{qr}$ capacitance (itself coupled to a transmission line thanks to a $C_\mathrm{c}$ capacitance), and to a drive line (via a $C_\mathrm{d}$ capacitance) fed with both an AC and DC bias $V_\mathrm{j}$ provided by a cryogenic bias-tee. The peculiarity of the circuits is the gate control ($V_\mathrm{g}$) of the Josephson junctions, thanks to a $C_\mathrm{g}$ capacitance: no magnetic field is used to control the critical current here. The primary role of $V_\mathrm{g}$ will be to control the Josephson energy ($E_\mathrm{J}$) by tuning the critical current of the graphene-based junction \cite{Heersche2007, Calado2015, Jan2018}. We address two devices, thereafter referred to as Q$_\mathrm{1}$ and Q$_\mathrm{2}$, which differ by the values of the capacitances (estimated using Ansys Maxwell electrostatic and Sonnet electromagnetic simulations), such that the quality factor ($Q_\mathrm{c}$) and charging energy ($E_\mathrm{C}$) are respectively larger ($1.30 \times 10^4$ vs $4.09\times10^3$) and smaller (343 MHz vs 516 MHz) for Q$_\mathrm{1}$ than for Q$_\mathrm{2}$ (see Table \ref{tab:capa}). Note that the charge dispersion, which we will discuss in details, is highly dependent on the charging energy. The resonant frequency $f_\mathrm{r}$ of the readout resonator is similar for the two devices, 6.95 GHz (Q$_\mathrm{1}$) and 6.79 GHz (Q$_\mathrm{2}$).

Device fabrication comprises four main steps (eight individual steps in total, see Appendix \ref{fab}). First, we define the aluminum superconducting circuit (transmission line, resonator, qubit, drive line, and gate) via e-beam lithography of a thin metal film (80 nm) deposited onto a high-resistivity silicon substrate. Second, we assemble the graphene weak link. To do so, we encapsulate graphene between two hexagonal boron nitride (\textit{h}-BN) layers ($\sim$~30~nm-thick each) using a dry transfer technique \cite{Pizzocchero2016}, to minimize the amount of airborne species in contact with graphene, which otherwise would harm the devices' performance. Third, graphene is side-contacted to Al electrodes (80 nm), after a buffer Ti layer (5 nm) has been deposited to improve the transmission of the metal/graphene contacts \cite{Wang2013}. The (graphene) weak link is 350 nm-wide and 350 nm-long. The fourth step is the fabrication of the topgate. For Q$_\mathrm{1}$, we use an extra layer of \textit{h}-BN to prevent any contact with the side electrodes, above which a 1 $\mu$m-wide Al electrode is defined by e-beam lithography followed by e-beam evaporation (80 nm). The topgate here spans the whole Josephson junction [Fig. \ref{fig:device}(b,c)]. On the contrary, for Q$_\mathrm{2}$, we used no extra \textit{h}-BN layer but a narrow, 100 nm-wide finger gate directly deposited on top of the \textit{h}-BN/graphene/\textit{h}-BN heterostructure [Fig. \ref{fig:device}(d)]. This requires an accurate alignment in e-beam lithography.

\begin{table}[b]
\caption{\label{tab:capa} Capacitances, coupling quality factor, and charging energy calculated for the two devices Q$_\mathrm{1}$ and Q$_\mathrm{2}$ from electrostatic and electromagnetic simulations.}
\begin{ruledtabular}
\begin{tabular}{cccccccc}
\textrm{}&
\textrm{$C_\mathrm{c}$}&
\textrm{$C_\mathrm{qr}$}&
\textrm{$C_\mathrm{S}$}&
\textrm{$C_\mathrm{d}$}&
\textrm{$C_\mathrm{g}$}&
\textrm{$Q_\mathrm{c}$}&
\textrm{$E_\mathrm{C}/h$}\\
\textrm{}&
\textrm{(fF)}&
\textrm{(fF)}&
\textrm{(fF)}&
\textrm{(aF)}&
\textrm{(fF)}&
\textrm{}&
\textrm{(MHz)}\\
\colrule
Q$_\mathrm{1}$ & $3.80$ & $4.46$ & $50.7$ & $176$ & $1.03$ & $1.30 \times 10^4$ & $343$\\
Q$_\mathrm{2}$ & $12.7$ & $3.71$ & $32.5$ & $168$ & $1.14$ & $4.09 \times 10^3$ & $516$
\end{tabular}
\end{ruledtabular}
\end{table}

\newpage

\section{\label{sec:modelsec}Graphene Josephson junction and circuit model\protect}

\begin{figure}[!htb]
\includegraphics{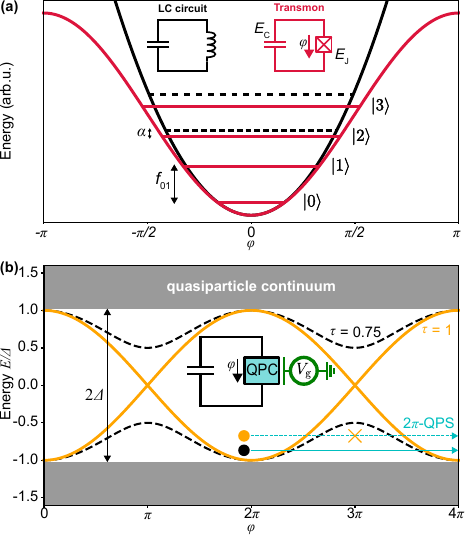}
\caption{\label{fig:model} Superconducting quantum point contact (S-QPC) model for a qubit based on a gate-tunable weak link between two superconducting electrodes. (a) Energy potential of the quantum harmonic oscillator (black) and the transmon qubit (red). In the latter, the Josephson junction's nonlinearity turns the energy spectrum anharmonic, such that energy levels are not evenly spaced. (b) Josephson potential of a S-QPC. When the transmission $\tau$ is below unity (black dotted line), $2 \pi$-quantum phase slips ($2 \pi$-QPSs) can occur, which is directly responsible for charge dispersion. For $\tau=1$ (solid orange line), the two ABS branches cross and $2 \pi$-QPS is forbidden, leading to a vanishing charge dispersion. Inset: effective model describing our devices as a gate-tunable S-QPC shunted by a capacitor.}
\end{figure}

Figure \ref{fig:model}(a) shows the energy levels within the parabolic potential of the quantum harmonic oscillator (black), and those within the potential of the so-called transmon qubit (red), in which the linear inductor is replaced by a nonlinear one, a Josephson junction. The latter spectrum is characterized by its anharmonicity, $\alpha$, defined as $f_\mathrm{12} = f_\mathrm{01} + \alpha$. 
In tunnel Josephson junctions, the coherent tunneling of Cooper pairs is well described by the traditional Josephson Hamiltonian $\hat{H}_\mathrm{TJ}=- E_\mathrm{J} \cos \left ( \hat{\varphi} \right )$, where $E_\mathrm{J}$ is the junction's Josephson energy and $ \hat{\varphi}$ the phase operator. When the weak link is not a tunnel barrier, for a semiconductor, a metal or a semi-metal (graphene), transmission of Cooper pairs from one electrode to the other is instead described using the concept of Andreev bound states (ABSs), involving electron-hole states within the weak link \cite{PhysRevLett.124.226801,PRXQuantum.5.030357,PhysRevLett.132.226301,PhysRevResearch.7.013248,PhysRevB.107.L140503,Ciaccia2024-fz,PhysRevB.108.094514}. The charge carrier density is hence a crucial parameter that will determine the number of channels contributing to the supercurrent, some of which can have high transmission ($\hat{H}_\mathrm{TJ}$ accounts only for low-probability tunneling processes). Several models have been used to describe highly transmitting junctions with a small number of channels. In the so-called short junction limit of a single ABS per conduction channel (i.e. when the length $L$ is small compared to the superconducting coherence length in the electrodes $\xi$), the energy of a single ABS is given by \cite{Kulik, Beenaker1991ABS, Titov2006, Bretheau2017}

\begin{equation}
E_{\mathrm{\pm}} = \pm \Delta \sqrt{1 - \tau \sin \left ( \frac{\varphi}{2} \right )^2}
\end{equation}

\noindent where '$\pm$' denotes the two branches of the ABS pair, $\Delta$ is the gap of the superconducting electrodes, and $\tau$ is the channel transmission. Assuming that only the ground energy ABS contributes, leads to the Josephson Hamiltonian $\hat{H}_\mathrm{ABS^\mathrm{-}}=-\Delta \sqrt{1 - \tau \sin \left ( \frac{\hat{\varphi}}{2} \right )^2}$, which approximates well to the tunnel Hamiltonian $\hat{H}_\mathrm{TJ}$ in the limit $\tau \ll 1$, for $E_\mathrm{J}=\frac{\Delta \tau}{4}$ \cite{Kringhøj2018,Zheng2023}. This model, limited to the ABS ground state, has been used to describe gatemon qubits based on III-V nanowires, wherein transmission through resonant levels indeed can lead to high transmission \cite{Kringhøj2018, De_lange}. Here, we propose a model to describe the complete ABS-pair.

The Josephson potential of a quantum point contact at perfect (solid orange line) and intermediate (dashed black line) transmission taking into account the ABSs is represented in Fig. \ref{fig:model}(b). The two ABS branches across the superconducting gap are only well-separated for small $\tau$ values. In other words, when transmission approaches unity, a proper description of the system must include both branches, as Landau-Zener like transitions between the ground and excited ABSs, are susceptible to occur where the two branches are the closest, i.e. $\varphi = \pi \; [2 \pi]$ \cite{Vakhtel2022}. Importantly, doing so to model superconducting quantum point contacts (S-QPC) \cite{Bargerbos2020,Kringhøj2020} or resonant tunneling \cite{Vakhtel2022}, one predicts a strong reduction of the charge dispersion, in case of a high transmission, owing to the suppression of $2 \pi$-quantum phase slips ($2 \pi$-QPSs). 
To describe our graphene-based junctions, we use as an effective model the Hamiltonian derived for a S-QPC \cite{Feigelman1999}, $\hat{H}_\mathrm{QPC}$ (see Appendix \ref{Hamiltonian}), which takes into account both ABS branches. The full Hamiltonian describing our circuit for a single conducting channel thus reads:

\begin{equation}
\hat{H} = 4 E_\mathrm{C} \left ( \hat{n} - n_\mathrm{g} \right ) ^2 + \Delta 
\left ( 
\begin{array}{cc}
    \cos \left ( \frac{\hat{\varphi}}{2} \right ) & \sqrt{r} \sin \left ( \frac{\hat{\varphi}}{2} \right ) \\
    \sqrt{r} \sin \left ( \frac{\hat{\varphi}}{2} \right ) & - \cos \left ( \frac{\hat{\varphi}}{2} \right )
    \end{array}
\right ).
\label{ITLZ H}
\end{equation}

Here, $\hat{n}$ and $\hat{\varphi}$ are the (conjugated) operators representing the number of Cooper pairs within the superconducting island [red portion of the circuit in Fig. \ref{fig:device}(a)] and the phase across the junction; $n_\mathrm{g}$ is the island’s charge offset. The $2 \times 2$ matrix encodes the contribution of both the ground and excited ABSs, which are coupled by backscattering, with $r=1-\tau$ and $\tau$ the channel transmission. In a 2D graphene weak link, however, there is a priori a certain number of conduction channels, $N_\mathrm{C}$ (that depends on $V_\mathrm{g}$, see Appendix \ref{Nc Vg}), each having a given transmission ($V_\mathrm{g}$-dependent too). This seems at odds with the QPC picture, yet the contributions of the few highest-transmission conduction channels are expected to dominate the Josephson energy $E_\mathrm{J}$ of the graphene-based junction. To limit the number of unknown parameters, we thus model the junction by $N_\mathrm{C} \left ( V_\mathrm{g} \right )$ independent ABSs of transmission $\tau$ (that depends on $V_\mathrm{g}$), which is equivalent to $N_\mathrm{C} \left ( V_\mathrm{g} \right )$ independent QPCs in parallel. This should be seen as an effective model not meant to describe the details of the junction, in particular its precise geometry. In this picture, the junction has consequently a gate-dependent Josephson energy $E_\mathrm{J} \left ( V_\mathrm{g} \right ) = \frac{\Delta N_\mathrm{C} \left ( V_\mathrm{g} \right ) \tau\left ( V_\mathrm{g} \right )}{4}$.
With these in mind, we numerically solve Hamiltonian \ref{ITLZ H} in the charge basis (rather than the phase basis, as it is usually done \cite{Kringhøj2020}), as detailled in Appendix \ref{deriv model}, to compute the energy spectrum and extract relevant experimental parameters such as charge dispersion and anharmonicity. The method for extracting the model parameters ($E_\mathrm{J}$, $\tau$) from the measured data is presented in Appendix \ref{lookup}.

\section{\label{sec:expsec}Frequency-tunability in a graphene superconducting qubit\protect}

We first use a standard two-tone scheme in continuous mode to measure the frequency of the first qubit transition, $f_\mathrm{01}$, as a function of the gate voltage. As discussed, the gate has mainly two effects on the graphene junction. First, it changes the Fermi wavelength, thereby opening new conduction channels. The second effect is a change of the channels' transmission. Experimentally, this is apparent as oscillations in the normal state resistance, $R_\mathrm{N}$, of the junction and is referred to as electronic Fabry-Pérot interferences \cite{Calado2015}. This effect is found in high quality devices hosting ballistic carrier transport. 

\begin{figure}[!htb]
\includegraphics{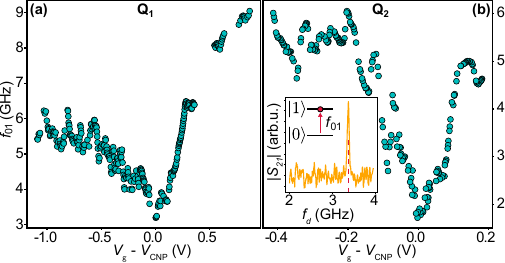}
\caption{\label{fig:energy-tunability} Gate-tunability of the qubit frequency. (a) Qubit frequency, $f_{\mathrm{01}}$, as a function of the gate voltage, $V_{\mathrm{g}}$ for device Q$_\mathrm{1}$. A pronounced asymmetry is observed about the charge neutrality point (CNP) of graphene ($V_{\mathrm{CNP}} \sim -1.81$ V). Reproducible oscillations are observed in the $p$-doped regime of graphene, which are attributed to a Fabry-Pérot effect. (b) Same kind of data for device Q$_\mathrm{2}$. The inset shows a typical two-tone measurement of the qubit dispersively coupled to a readout resonator. When the drive tone frequency, $f_{\mathrm{d}}$, matches the qubit frequency, the qubit is excited from state $\ket{0}$ to state $\ket{1}$. The resonant frequency of the resonator is shifted accordingly and probed by a change in the amplitude, $|S_{\mathrm{21}}|$, of the readout tone.}
\end{figure}

The $f_\mathrm{01} \left ( V_\mathrm{g} \right )$ data for devices Q$_\mathrm{1}$ and Q$_\mathrm{2}$ are shown in Figs. \ref{fig:energy-tunability}(a,b), respectively. When $V_\mathrm{g}$ hits the crossover between the hole- and electron-conduction regimes in graphene, i.e. at the charge neutrality point (CNP), the critical current reaches a minimal value, resulting in maximal Josephson inductance $L_\mathrm{J}$ and minimal $f_\mathrm{01}$. Moving away from the CNP, the transition frequency increases in both devices. In device Q$_\mathrm{1}$, a strong asymmetry in $f_\mathrm{01}$ with respect to $V_\mathrm{CNP}$ is observed. For $V_\mathrm{g} > V_\mathrm{CNP}$ ($n$-doped graphene), the qubit frequency increases quickly with charge carrier density, from $\sim 3$ to $\sim 9$ GHz within $< 1$ V. Such large frequency tuning in a narrow gate voltage range is appealing for applications that require fast frequency tuning of the transition, such as tunable coupling \cite{Steffen2005} or Z-gate \cite{Steffen2006}. For $V_\mathrm{g} < V_\mathrm{CNP}$ ($p$-doped graphene), $f_\mathrm{01}$ increases more progressively while oscillating. Similar behaviour was reported in Ref. \cite{Jan2018} and is reminiscent of what is observed when measuring the critical current of such graphene Josephson junctions \cite{Park2018} (see also Appendix \ref{meas}). The asymmetry is attributed to the electron doping induced by the Ti/Al contacts, which decreases the transmission at the interfaces in the $p$-doped regime of graphene because of the presence of a $pn$ junction \cite{PhysRevLett.101.026803, Calado2015}.

In device Q$_\mathrm{2}$, we also observe a large frequency tunability (several GHz) and oscillations but no clear asymmetry between the $p$-doped and $n$-doped regimes. We attribute this difference to the gate geometry, which is different from that in Q$_\mathrm{1}$. As illustrated in Fig.~\ref{fig:device}(d), in device Q$_\mathrm{2}$ the narrow gate only partly covers the junction, presumably resulting in a more complex electrostatic landscape, and charge density profile, than in Q$_\mathrm{1}$.

\section{\label{sec:expsec2}Tunable anharmonicity and charge dispersion\protect}

We now focus on two key characteristics of the energy spectrum of a qubit, namely its anharmonicity $\alpha$, and the dispersion of the first qubit transition with the charge offset, $f_\mathrm{01} \left ( n_\mathrm{g} \right )$. Using a top-gated graphene weak link ($V_\mathrm{g}$) and a gate voltage on the superconducting island of the qubit ($V_\mathrm{j}$), both quantities can be tuned. Below we analyze the dependence on the gate voltage $V_\mathrm{g}$, or rather on the Josephson energy, which is a function of $V_\mathrm{g}$ and is a main parameter in the model. While the charge dispersion, i.e. $\delta f_\mathrm{01} = f_\mathrm{01} \left ( n_\mathrm{g}=0 \right ) - f_\mathrm{01} \left ( n_\mathrm{g}=0.5 \right )$, is too small to be measured in device Q$_\mathrm{1}$, owing to the large $E_\mathrm{J}/E_\mathrm{C}$ in this device ($>$ 15), it becomes appreciable in device Q$_\mathrm{2}$, whose $E_\mathrm{J}/E_\mathrm{C}$ ratio can be set in a lower range (2 to 4) due to the larger designed $E_\mathrm{C}$. Although this is certainly antinomic with a long-lasting coherence (hence, not a desirable property for a qubit), with this device we are able to show that $\delta f_\mathrm{01}$ can be reduced, not only by circuit geometry (i.e. by setting an $E_\mathrm{J}/E_\mathrm{C}$ ratio by design), but via the action of the topgate on graphene.

\begin{figure}[!htb]
\includegraphics{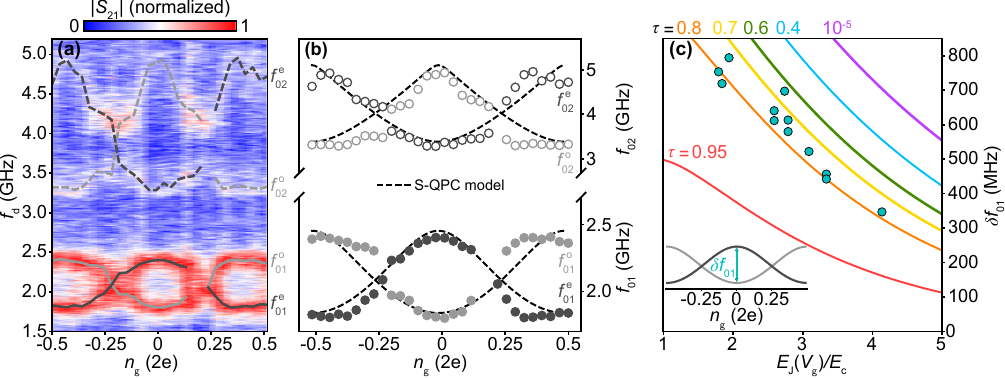}
\caption{\label{fig:dispersion} Qubit spectrum and charge dispersion in device Q$_\mathrm{2}$. (a) Two-tone measurement of the first two qubit transitions as a function of the charge offset, $n_{\mathrm{g}}$, at $V_{\mathrm{g}} = 0.41$ V. Solid and dashed lines represent the fitted frequency of the two charge parities (odd and even, labeled 'o' and 'e') of the $\ket{0}$-$\ket{1}$ and $\ket{0}$-$\ket{2}$ transitions. (b) Reconstruction of the qubit spectrum at $V_\mathrm{g}=0.41$~V using only the set of $\left (\delta f_{\mathrm{01}}, f_{\mathrm{01}} \right )$ values derived from the experimental data. The model (dashed lines) accounts for the $\ket{0}$-$\ket{1}$ transition, and correctly reproduces the $\ket{0}$-$\ket{2}$ transition too, without free parameters. (c) Charge dispersion, $\delta f_{\mathrm{01}}$, as a function of $E_\mathrm{J} \left ( V_\mathrm{g} \right ) / E_\mathrm{C}$, extracted from data such as in (a) (as shown in the inset), varying  $V_{\mathrm{g}}$ from 400 mV to 486 mV. The coloured solid lines represent the charge dispersion predicted by Hamiltonian \ref{ITLZ H} for different values of transmission.}
\end{figure}

Using a large drive power, we observe the first two transitions, with frequency $f_\mathrm{01}$ and $f_\mathrm{02}$ when varying the drive frequency $f_\mathrm{d}$ in a two-tone measurements. We will at first not discuss  the second excited state but we will come back to it later when dealing with the anharmonicity. Figure \ref{fig:dispersion}(a) presents the result of the two-tone measurement, as a function of $n_\mathrm{g}$, for $V_\mathrm{g}=0.41$ V, from which we deduce a $\delta f_\mathrm{01}$ value of $614 \pm 4$ MHz. The energy levels appear in the form of two branches, each corresponding to a different parity of the charge states. This is due to quasiparticle poisonning, occuring at much faster time-scale than the few-seconds duration of a full sweep of $f_\mathrm{d}$ (specific designs, beyond the scope of the present work, exist to reduce poisonning \cite{Pan2022}).

To study the effect of the gate voltage, we select a low $E_\mathrm{J}/E_\mathrm{C}$ near the charge neutrality point of device Q$_\mathrm{2}$ (within a gate voltage range of about 100 mV). Varying $V_\mathrm{g}$ in this range, we find that $\delta f_\mathrm{01}$ values span between $\sim 800$ and $\sim 350$~MHz. This is represented in Fig.~\ref{fig:dispersion}(c) (cyan data points) as a function of $E_\mathrm{J}/E_\mathrm{C}$, which increases by a factor two with $V_\mathrm{g}$, while the transmission stays in the range 0.6 to 0.8. Here, the extracted $E_\mathrm{J}$ varies from about 1 to 2 GHz. Such values could be compared to the one expected for a single fully transmitting channel between aluminum electrodes. From the bulk aluminum superconducting gap of  $\Delta_\mathrm{Al} \sim 200$ $\mu$eV, one would expect $E_\mathrm{J} \sim$ 10 GHz in this case. In this model, one would reach the unphysical conclusion that in this regime, close to the Dirac point, $N_\mathrm{C} < 1$. This shows the limitation of the model close to the neutrality point. In particular, the assumption of an energy-independent transmission made in the S-QPC model might not be valid here. While this can be an appropriate assumption for a metal, or for highly-doped graphene ($E_\mathrm{F} \gg \Delta$), we see that this does not hold near the neutrality point. Another model, such as a resonant-level one with an energy-dependent transmission may be more relevant to describe the junction at very low doping \cite{PhysRevB.104.174517}.

In Fig.~\ref{fig:dispersion}, we also plotted the predictions from the effective Hamiltonian of Eq.~\ref{ITLZ H} for different values of the transmission $\tau$. According to our model, the experimental datapoint correspond to an effective transmission of the junction $\tau$ comprised between 0.6 and 0.8. This is consistent with what has been derived from measurements of the current-phase relation (CPR) of graphene Josephson junctions \cite{Lee2015-pi,jha2024large,Messelot} and tunnel spectroscopy of the Andreev levels dispersion \cite{Bretheau2017}. Importantly, our model indicates that the large transmission values, compared to a tunnel junction, translate into a substantial reduction of the charge dispersion, as illustrated in Fig. \ref{fig:dispersion}(c) (and further discussed in Appendix \ref{deriv model}). This trend is not restricted to graphene, and also applies in gatemon qubits based on InAs nanowires weak links \cite{Bargerbos2020,Kringhøj2020}. In a charge qubit with a relatively small $E_\mathrm{J}/E_\mathrm{C}$ ratio (i.e. away from the transmon regime), as is the case of our Q$_\mathrm{2}$ device, this could be used to mitigate the sensitivity to charge noise while preserving a large anharmonicity ($\alpha \sim -1.5$ GHz here for $n_\mathrm{g}=0$), the latter being required for fast control of the qubit \cite{10.1063/1.5089550}. In our top-gated graphene-based qubit, the channels' transmission can further be increased with $V_\mathrm{g}$, and the charge dispersion reduced accordingly, as also shown in Fig. \ref{fig:dispersion}(c).

To further test the robustness and consistency of our model, especially in the vicinity of the charge neutrality point, we come back to the measurement of the frequencies of the first two excited states as a function of the charge offset, $f_\mathrm{01}$ and $f_\mathrm{02}$, in device Q$_\mathrm{2}$ at $V_\mathrm{g}=0.41$ V [disk symbols in Fig. \ref{fig:dispersion}(b)]. From Fig. \ref{fig:dispersion}(c) (i.e. from the experimental qubit first transition frequency $f_\mathrm{01}$ and charge dispersion $\delta f_\mathrm{01}$), our model predicts an effective transmission $\tau = 0.73$ and $E_\mathrm{J}/E_\mathrm{C} = 2.79$. We use these values to derive $f_\mathrm{02} \left ( n_\mathrm{g} \right )$ according to Hamiltonian \ref{ITLZ H} (no additional parameter needs to be introduced here) [black dashed lines in Fig. \ref{fig:dispersion}(b)]. The agreement with the data points is very good (the $\ket{0}$-$\ket{2}$ transition is well reconstructed), further confirming that our effective model, in the short junction limit, describes well our system.

\begin{figure}[!htb]
\includegraphics{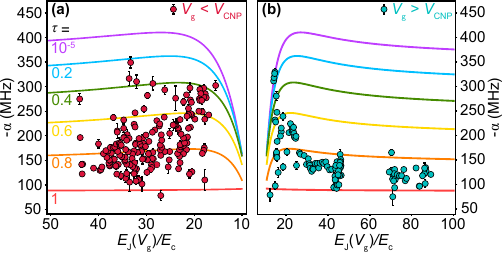}
\caption{\label{fig:anharmonicity} Gate-dependent anharmonicity in device Q$_\mathrm{1}$. (a) Anharmonicity, $-\alpha$, as a function of $E_\mathrm{J} \left ( V_\mathrm{g} \right ) / E_\mathrm{C}$, where $E_{\mathrm{J}}$ is tuned by the gate voltage, in the $p$-doped conduction regime of graphene. The solid lines are the model predictions for a set of transmissions. (b) Similar data in the $n$-doped regime of graphene.}
\end{figure}

We finally turn to the qubit's anharmonicity $\alpha$ in device Q$_\mathrm{1}$, which we experimentally determine as a function of $V_\mathrm{g}$ by comparing $f_\mathrm{01}$ and $f_\mathrm{02}$ (here obtained by a two-photon process at large power \cite{Kringhøj2018}), i.e. $\alpha = 2 \left ( \frac{f_\mathrm{02}}{2} - f_\mathrm{01} \right )$. The corresponding data are shown in Appendix \ref{meas}, and are represented as a function of the inferred $E_\mathrm{J} \left ( V_\mathrm{g} \right ) /E_\mathrm{C}$ ratio using our model in Fig. \ref{fig:anharmonicity}(a,b) in the $n$-doped and $p$-doped regimes of graphene, respectively. We first note that $-\alpha$ remains small, never exceeding 350 MHz, which is close to the charging energy of the device ($E_\mathrm{C}/h = 343$ MHz). Values lower than $E_\mathrm{C}$  are not unusual with gatemon qubits, generally operated in the transmon regime, and signal a close-to-harmonic Josephson potential tied to a high channel transmission \cite{Kringhøj2018, Hertel, Zheng2023, Zhuo2023, Sagi2024, Kiyooka2024}. Away from the low-transmission limit, the nonzero-length ($L$) weak link should additionally further reduce the anharmonicity \cite{Fatemi2024}. The (small) $-\alpha$ presents strong relative variations, reaching values below 100 MHz. For $V_\mathrm{g} < V_\mathrm{CNP}$, i.e. in the $p$-doped conduction regime, Fig. \ref{fig:anharmonicity}(a) reveals large reproducible anharmonicity fluctuations, which seem to be correlated to those in $f_\mathrm{01}$ [Fig. \ref{fig:energy-tunability}(a)]. According to our model, these fluctuations can be tracked back to large variations of the effective transmission of the junction with $V_\mathrm{g}$, from 0.2 to 1. In contrast, in the $n$-doped conduction regime in graphene [Fig. \ref{fig:anharmonicity}(b)], $\alpha$ exhibits less fluctuations, with a global tendency to decrease with $E_\mathrm{J} \left ( V_\mathrm{g} \right )$ towards a small fraction of the charging energy. At this limit, our model predicts a transmission close to unity. Overall, covering a broad range of $E_\mathrm{J}/E_\mathrm{C}$ values (15 to 90) through a $V_\mathrm{g}$ excursion of less than 1 V, we demonstrate a tuning of the qubit's anharmonicity from $-E_\mathrm{C}/h$ to $-E_\mathrm{C}/4h$. To retain a large enough anharmonicity while conserving gate-tunability, it has been shown that other designs, such as the flux-tunable transmon \cite{De_lange}, or the fluxonium \cite{PRXQuantum.6.010326}, could favorably integrate gate-tunable weak links. 

\section{\label{sec:conclusion}Conclusion\protect}

We have shown that using Josephson junctions with a top-gated graphene weak link, we can build charge qubits operating in a broad range of $E_\mathrm{J}/E_\mathrm{C}$ ratios and having highly gate-tunable energy spectrum. The fundamental qubit transition is sensitively tuned, by several GHz, via the topgate voltage. Designing a device with small $E_\mathrm{J}/E_\mathrm{C}$, which is intrinsically prone to significant charge dispersion, we show that a high effective transmission of the Josephson junction mitigates this charge dispersion, which can be further reduced by in-situ increasing $E_\mathrm{J}$, again via $V_\mathrm{g}$. The anharmonicity of our graphene-based qubits at large $E_\mathrm{J}/E_\mathrm{C}$ is relatively small, partly due to the high transmission of the junctions, but can be increased using a more complex design in the future \cite{PRXQuantum.6.010326,De_lange}. To rationalize our results and model a graphene superconducting qubit, we have used and discussed a model initially introduced to describe superconducting quantum point contacts, which takes into account the ground and excited states of the Andreev bound states in the weak link. The model has essentially two tunable key effective parameters, the Josephson energy and effective transmission of the conduction channels in the weak link. This effective model quantitatively describes our results, except close to the charge neutrality point where the assumption of an energy-independent transmission seems not to hold. A resonant-level model might be more appropriate to decsribe this specific regime of only a few conducting channels \cite{PhysRevB.104.174517}. Our work highlights graphene-based hybrid junctions as promising candidates for qubits with low noise susceptibility, which, together with the efficient gate-tunability, holds promises for advanced functionnalities in superconducting quantum circuits. A more robust control of the channels' transmission using local gates near the superconducting electrodes \cite{PhysRevX.8.031023,10.1063/1.5021113,PhysRevLett.120.057701,10.1063/5.0241328,PhysRevX.15.011046} towards a complete suppression of charge dispersion might further improve the quantum coherence of such circuits.

\section*{Data Availability Statement}

The supporting data for this article are openly available from the Zenodo repository with the identifier \url{https://doi.org/10.5281/zenodo.15511640}.

\begin{acknowledgments}

This work was supported by the French National Research Agency (ANR) in the framework of the Graphmon project (ANR-19-CE47-0007). This work benefited from a French government grant managed by the ANR agency under the ‘France 2030 plan’, with reference ANR-22-PETQ-0003. K.W. and T.T. acknowledge support from the JSPS KAKENHI (Grant Numbers 21H05233 and 23H02052) and World Premier International Research Center Initiative (WPI), MEXT, Japan. We acknowledge the staff of the Nanofab cleanroom of Institut Néel for help with device fabrication. We thank Richard Haettel for assistance in device annealing, Simon Le Denmat for atomic force microscopy and Aloïs Arrighi for sharing the PDMS transfer technique. We acknowledge the work of Julien Jarreau, Laurent Del-Rey and Didier Dufeu for the design and fabrication of the sample holders and other mechanical pieces used in the cryogenic system. We thank Denis M. Basko for fruitful discussions and help with numerical implementations of the models. We also thank Landry Bretheau, Sergey Frolov, Tereza Vakhtel for discussions.

\end{acknowledgments}

\appendix

\section{FABRICATION TECHNIQUES}
\label{fab}

We first exfoliated bulk graphite and \textit{h}-BN crystals on Si/SiO$_\mathrm{2}$ substrates. Relevant flakes are assembled accordingly to make a \textit{h}-BN/graphene/\textit{h}-BN heterostructure using a dry polymer pick-up technique with polypropylene carbonate (PPC) \cite{Zomer2014, Kim2016}. To prepare the substrate, we evaporate a 80 nm Al film on top of a high resistivity Si substrate. A first e-beam lithography step is performed to define alignement marks for subsequent e-beam lithographies.  We used a second e-beam lithography step to define most of the superconducting circuit: transmission line, resonator, qubit island, drive line and gate. After a standard MIBK-IPA development, the exposed Al is removed by dipping the chip in a H$_\mathrm{3}$PO$_\mathrm{4}$ + HNO$_\mathrm{3}$ solution for about 3 min and 30 s (the precise etching rate is calibrated each time with a control sample). Next, another e-beam lithography is made to open a square window of 100 $\mu$m lateral dimension in the PMMA film. This ensures a cleaner circuit after the stack deposition. The heterostructure is then deposited onto the substrate by contacting and melting the PPC at 145°C. The polymer and potential parasitic flakes are then washed away in an acetone bath overnight and any remaining PPC is removed by annealing the chip in a vacuum oven at 350°C for 2 hours. The Ti/Al contacts are defined using an e-beam lithography step. We used a cold development in a 3:1 IPA:DI water  solution at 3°C for 90 s. After a 60 s electrodionized (EDI) water rinse to stop the reaction, the top \textit{h}-BN, graphene and a portion of the bottom \textit{h}-BN are etched using reactive ion etching with a CHF$_\mathrm{3}$ + O$_\mathrm{2}$ mixture. Immediately after, we deposited the Ti/Al (5 nm/80 nm) superconducting electrodes using an e-beam evaporator to form edge-contacts separated by about 350~nm in a graphene flake \cite{Wang2013}. We then use a lift-off technique to remove excess metal in an acetone bath. The width of the junction is defined in a subsequent e-beam lithography step, in which extra graphene is etched away using reactive ion etching. For device Q$_\mathrm{1}$, the gate dielectric is exfoliated on a Gelpack 4 polydimethylsiloxane (PDMS) stamp and released on top of the junction by precise optical alignement and contact to the chip. The stamp is then heated up to 70°C and slowly lifted up. PDMS residues are eliminated using toluene at 80°C overnight. For device Q$_\mathrm{2}$, the top \textit{h}-BN layer directly serves as the gate dielectric and a narrow (100 nm-wide) finger gate is aligned in between the contacts using e-beam lithography. The 80 nm-thick Al gate electrode is deposited using e-beam evaporation. Finally, the contacts and gate electrode are connected to the rest of the superconducting circuit using a last step of e-beam lithography. We used in-situ argon milling to etch the aluminum natural oxide layer before evaporating the metal, i.e. 120~nm of aluminum.

\section{ESTIMATION OF THE GATE-DEPENDENT NUMBER OF CONDUCTIVE CHANNELS}
\label{Nc Vg}

\begin{figure}[!htb]
\includegraphics{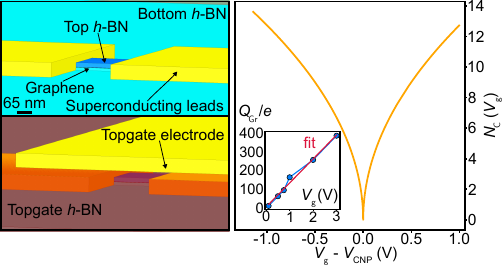}
\caption{\label{fig:NcVg} Estimation of the gate-dependent number of conductive channels in the weak link. (a) Simulated design of device Q$_\mathrm{1}$ in Ansys Maxwell. (top) The graphene-based Josephson junction with superconducting leads. (bottom) The topgate made of an additional \textit{h}-BN layer and the top superconducting electrode. (b) Number of conductive channels $N_\mathrm{C}$ as a function of the gate voltage $V_\mathrm{g}$. The inset shows the electrostatic simulation of the excess charge $Q_\mathrm{Gr}$ in graphene as a function of $V_\mathrm{g}$. A linear fit is made to extract the $C_\mathrm{Gr}$ capacitance.}
\end{figure}

The control of the charge carrier density in the graphene weak link, $n_\mathrm{Gr}$, via the gate voltage $V_\mathrm{g}$ is given by \cite{Bretheau2017}

\begin{equation}
\begin{aligned}
        n_\mathrm{Gr} \left ( V_\mathrm{g} \right ) = \frac{C_\mathrm{Gr} | V_\mathrm{g}-V_\mathrm{CNP}|}{Se}
\end{aligned}
\end{equation}
with $C_\mathrm{Gr}$ the capacitance between the gate line and the graphene sheet, $S=350$ nm$^2$ graphene's surface, and $e$ the electron's charge. This translates into a Fermi wavelength $\lambda_\mathrm{F} \left ( V_\mathrm{g} \right ) = 2 \sqrt{\frac{\pi S e}{C_\mathrm{Gr} |V_\mathrm{g}-V_\mathrm{CNP}|}}$, and the number of conduction channels, $N_\mathrm{C}$, can be estimated by $N_\mathrm{C} \left ( V_\mathrm{g} \right ) \sim 4 W/\lambda_\mathrm{F} \left ( V_\mathrm{g} \right )$, with $W$ the width of the graphene weak link. An accurate estimation of $C_\mathrm{Gr}$ allows to obtain $N_\mathrm{C}$. We use electrostatic simulations to compute the capacitance between the gate line and graphene in device Q$_\mathrm{1}$ with a topgate geometry [see Fig. \ref{fig:device}(b,c)], by considering its environment (especially the metallic superconducting electrodes, which can screen the gate's electric field) [Fig. \ref{fig:NcVg}(a)]. Graphene lies on top of a silicon substrate ($\epsilon_\mathrm{Si} =11.9$), and is encapsulated in between two \textit{h}-BN layers ($\epsilon_\mathrm{\textit{h}-BN} =3.76$ \cite{Laturia2018,Wang2021}). Inset of Fig. \ref{fig:NcVg}(b) shows the excess charge $Q_\mathrm{Gr}$ in graphene as a function of the gate voltage $V_\mathrm{g}$. We use a linear fit to extract the capacitance between the gate line and graphene, which yields $C_\mathrm{Gr} = 20.4 \pm 1.1$ aF. The number of conduction channels [Fig. \ref{fig:NcVg}(b)] is calculated for the same gate voltage range than the one used in device Q$_\mathrm{1}$. By sweeping the gate voltage by 1 V from the CNP in graphene, we expect $\sim 12$ additional conductive channels in our graphene-based Josephson junction.

\section{CHARGE BASIS DERIVATION OF THE MODEL}
\label{deriv model}

\subsection{Charge qubit Hamiltonian}

For clarity, we first reproduce here the derivation of the charge qubit model for a tunnel Josephson junction, i.e. $\hat{H}_\mathrm{charge} = 4 E_\mathrm{C} \left ( \hat{n} - n_\mathrm{g} \right )^2 - E_\mathrm{J} \cos \left ( \hat{\varphi} \right )$ in the charge basis, with $\cos \left ( \hat{\varphi} \right ) = \frac{e^{i \hat{\varphi}} + e^{-i \hat{\varphi}}}{2}$. The operators $e^\mathrm{\pm i \hat{\varphi}}$ are obtained from the conjugation relation between $\hat{\varphi}$ and $\hat{n}$, namely $\left [ \hat{\varphi}, \hat{n} \right ] = i$. Employing Lie algebra relation ($\left [ \hat{\varphi}^\mathrm{k}, \hat{n} \right ] = ik \hat{\varphi}^\mathrm{k-1}$ for $k \in \mathbb{N}$) and the exponential series for a complex number $z$, one gets

\begin{equation}
\begin{aligned}
    e^\mathrm{\pm i\hat{\varphi}} \times \hat{n} &= \left [\sum_\mathrm{k=0}^\mathrm{\infty} \frac{\left(\pm i\hat{\varphi}\right)^\mathrm{k}}{k!} \right ] \times \hat{n} \\
    &= \sum_\mathrm{k=0}^\mathrm{\infty} \frac{i^\mathrm{k} \left(\pm \hat{\varphi}\right)^\mathrm{k} \hat{n}}{k!} \\
     &= \sum_\mathrm{k=0}^\mathrm{\infty} \frac{i^\mathrm{k} \hat{n} \left(\pm \hat{\varphi}\right)^\mathrm{k}}{k!} + \sum_\mathrm{k=1}^\mathrm{\infty} \frac{i^\mathrm{k} i k \left(\pm \hat{\varphi}\right)^\mathrm{k-1}}{k!} \\   
     &= \hat{n} \times \sum_\mathrm{k=0}^\mathrm{\infty} \frac{\left(\pm i \hat{\varphi}\right)^\mathrm{k}}{k!} + i \sum_\mathrm{k=1}^\mathrm{\infty} \frac{ i k \left(\pm i \hat{\varphi}\right)^\mathrm{k-1}}{k!}
     \\
     &=\hat{n} \times e^\mathrm{\pm i \hat{\varphi}}  + i \frac{ d e^\mathrm{\pm i \hat{\varphi}}}{d \hat{\varphi}} \\
     &= \hat{n} \times e^\mathrm{\pm i \hat{\varphi}}  \mp e^\mathrm{\pm i \hat{\varphi}}
\end{aligned}    
\end{equation}

Such that 

\begin{equation}
\begin{aligned}
        \left[e^\mathrm{\pm i \hat{\varphi}}, \hat{n}\right] = \mp e^\mathrm{\pm i \hat{\varphi}}
\end{aligned}
\end{equation}

In the orthonormal charge basis, the operators $e^\mathrm{\pm i {\hat{\varphi}}}$ can then directly be written as

\begin{equation}
\begin{aligned}
        e^\mathrm{\pm i {\hat{\varphi}}} = \sum_\mathrm{k=-\infty}^\mathrm{\infty} \ket{k \mp 1} \bra{k}
\end{aligned}
\end{equation}

This corresponds to a charge translation operator. The Josephson potential can thus be obtained 

\begin{equation}
    \begin{aligned}
        E_\mathrm{J} \cos \left ( \hat{\varphi} \right ) = \frac{E_\mathrm{J}}{2} \sum_\mathrm{k} \left [ \ket{k-1} \bra{k} + \ket{k+1} \bra{k} \right ]
    \end{aligned}
\end{equation}

In the charge basis $\left \{ \ket{k} \right \}$, the charge qubit Hamiltonian for a tunnel Josephson junction is accordingly

\begin{equation}
    \begin{aligned}
        \hat{H}_\mathrm{charge} = \left [\sum_\mathrm{k} 4 E_\mathrm{C} \left ( n_\mathrm{k} - n_\mathrm{g} \right )^2 \ket{k} \bra{k} - \frac{E_\mathrm{J}}{2} \left ( \ket{k-1} \bra{k} + \ket{k+1} \bra{k} \right ) \right ] 
    \end{aligned}
\end{equation}

\subsection{Superconducting QPC Hamiltonian}

We now detail the derivation of Eq. \ref{ITLZ H} in the main text, in the charge basis using what we derived previously. The Josephson Hamiltonian can be rewritten as

\begin{equation}
    H_\mathrm{QPC} \left ( \hat{\varphi} \right ) = \Delta \begin{pmatrix}
        \frac{ e^{\frac{i \hat{\varphi}}{2}} + e^{-\frac{i \hat{\varphi}}{2}} }{2}&\sqrt{r} \frac{ e^{\frac{i \hat{\varphi}}{2}} - e^{-\frac{i \hat{\varphi}}{2}} }{2i}\\
        \sqrt{r} \frac{ e^{\frac{i \hat{\varphi}}{2}} - e^{-\frac{i \hat{\varphi}}{2}} }{2i}&-\frac{ e^{\frac{i \hat{\varphi}}{2}} + e^{-\frac{i \hat{\varphi}}{2}} }{2}
    \end{pmatrix}
    \label{Josephson potential ITLZ}
\end{equation}

The operators $e^\mathrm{\pm i \frac{\hat{\varphi}}{2}}$ are similarly obtained by the use of $\left [ \frac{\hat{\varphi}}{2}, \hat{n} \right ] = \frac{i}{2}$

\begin{equation}
\begin{aligned}
    e^\mathrm{\pm i\frac{\hat{\varphi}}{2}} \times \hat{n} &= \left [\sum_\mathrm{k=0}^\mathrm{\infty} \frac{\left(\pm i\frac{\hat{\varphi}}{2}\right)^\mathrm{k}}{k!} \right ] \times \hat{n} \\
    &= \sum_\mathrm{k=0}^\mathrm{\infty}  \frac{\left (\frac{i}{2} \right )^\mathrm{k} \left(\pm \hat{\varphi}\right)^\mathrm{k} \hat{n}}{k!} \\
    &= \sum_\mathrm{k=0}^\mathrm{\infty}  \frac{\left (\frac{i}{2} \right )^\mathrm{k} \hat{n} \left(\pm \hat{\varphi}\right)^\mathrm{k}}{k!} + \sum_\mathrm{k=1}^\mathrm{\infty}  \frac{\left (\frac{i}{2} \right )^\mathrm{k} i k \left(\pm \hat{\varphi}\right)^\mathrm{k-1}}{k!} \\
     &= \hat{n} \times  \sum_\mathrm{k=0}^\mathrm{\infty} \frac{\left(\pm i\frac{\hat{\varphi}}{2}\right)^\mathrm{k}}{k!} + i \sum_\mathrm{k=1}^\mathrm{\infty} \frac{i k \left(\pm  \frac{\hat{\varphi}}{2}\right)^\mathrm{k-1}}{2k!} \\   
     &=\hat{n} \times e^\mathrm{\pm i \frac{\hat{\varphi}}{2}}  + i \frac{ d e^\mathrm{\pm i \frac{\hat{\varphi}}{2}}}{d \hat{\varphi}} \\
     &= \hat{n} \times e^\mathrm{\pm i \frac{\hat{\varphi}}{2}}  \mp \frac{1}{2} e^\mathrm{\pm i \frac{\hat{\varphi}}{2}}
\end{aligned}    
\end{equation}

Such that 

\begin{equation}
\begin{aligned}
        \left[e^\mathrm{\pm i \frac{\hat{\varphi}}{2}}, \hat{n}\right] = \mp \frac{1}{2} e^\mathrm{\pm i \frac{\hat{\varphi}}{2}}
\end{aligned}
\end{equation}

In the orthonormal charge basis, the operators $e^\mathrm{\pm i {\frac{\hat{\varphi}}{2}}}$ can then directly be written as

\begin{equation}
\begin{aligned}
        e^\mathrm{\pm i {\frac{\hat{\varphi}}{2}}} = \sum_\mathrm{k=-\infty}^\mathrm{\infty} \ket{k \mp \frac{1}{2}} \bra{k}
\end{aligned}
\end{equation}

The $2 \times 2$ subnetwork implementing the upper and lower ABSs is numerically treated in the form of Kronecker products. First, we built the kinetic energy term by using a Kronecker product between the charging energy term and the $2 \times 2$ identity matrix. The diagonal and off-diagonal terms proportionnal to $e^\mathrm{\pm i {\frac{\hat{\varphi}}{2}}}$ in the Josephson potential [Eq. \ref{Josephson potential ITLZ}] are then dealt with by using a Kronecker product between $e^\mathrm{\pm i {\frac{\hat{\varphi}}{2}}}$ and the $z$-Pauli matrix, $\sqrt{r}e^\mathrm{\pm i {\frac{\hat{\varphi}}{2}}}$ and the $x$-Pauli matrix, respectively. The code is available from the Zenodo repository.

\subsection{Calculation of the charge dispersion}

\begin{figure}[!htb]
\includegraphics{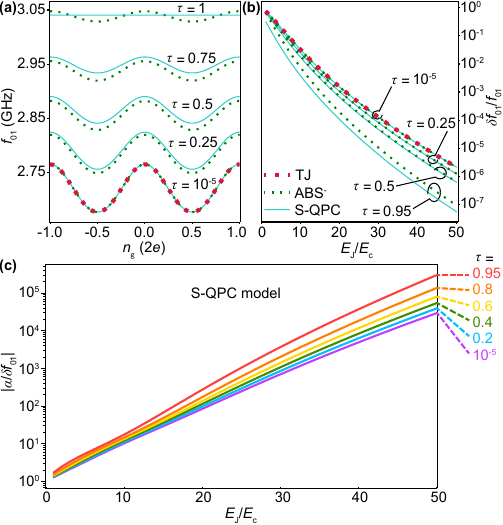}
\caption{\label{fig:simu charge disp} Numerical simulation of the qubit spectrum in the tunnel junction limit, for a model only taking into account the lower energy ABS, and the superconducting quantum point contact model. (a) Qubit transition frequency, $f_\mathrm{01}$, as a function of $n_\mathrm{g}$ for $E_\mathrm{J}/E_\mathrm{C} = 10$. A larger transmission increases the qubit frequency and decreases the amplitude of the $f_\mathrm{01} \left ( n_\mathrm{g} \right )$ oscillations. (b) Normalized charged dispersion, $\delta f_\mathrm{01} / f_\mathrm{01}$, suppressed as a function of $E_\mathrm{J}/E_\mathrm{C}$, which is expected in the transmon regime. An additional suppression of the charge dispersion is predicted by a large transmission. (c) $|\alpha / \delta f_\mathrm{01}|$ vs $E_\mathrm{J}/E_\mathrm{C}$ for different transmissions obtained with the S-QPC model for $n_\mathrm{g}=0$. This shows the benefit of using a high-transmission junction to reduce charge dispersion despite having a more harmonic potential.}
\end{figure}

We calculated the charge dispersion based on the Hamiltonian of a superconducting charge qubit using the 3 different Josephson potentials Hamiltonians discussed in the main text (i.e. $\hat{H}_\mathrm{TJ}$, $\hat{H}_\mathrm{ABS^\mathrm{-}}$, and $\hat{H}_\mathrm{QPC}$) to model the junction, as a function of $E_\mathrm{J}/E_\mathrm{C}$ and $\tau$. To do so, we truncated the Hilbert space at 201 states (i.e. a charge basis $\left \{ \ket{-100}, \ldots, \ket{100} \right \}$) and calculated the circuit eigenenergies for both $n_\mathrm{g}=0$ and $n_\mathrm{g}=0.5$ and computed the charge dispersion, defined as $\delta f_\mathrm{01} = f_\mathrm{01} \left ( n_\mathrm{g}=0 \right ) - f_\mathrm{01} \left ( n_\mathrm{g}=0.5 \right )$. We present the results obtained with $E_\mathrm{C}/h = 350$ MHz in Fig.~\ref{fig:simu charge disp}. We observe that the charge dispersion is accordingly suppressed with $E_\mathrm{J}/E_\mathrm{C}$ (as shown in Ref. \cite{Koch2007}) but also with the transmission. Using $\hat{H}_\mathrm{ABS^\mathrm{-}}$, derived in the phase basis, the Josephson potential shape and height are modified with $\tau$, which indicates that the tunneling amplitude of coherent quantum phase slips between neighboring Josephson potential minima is reduced \cite{Vakhtel2022}. By switching to $\hat{H}_\mathrm{QPC}$ and adding the excited state of the ABS in the picture, the charge dispersion at a given transmission is further reduced. This reduction is even greater when the transmission is high. This is an indication that the upper energy ABS indeed starts to play a role in the system, as discussed in the main text.

\section{COMMENT ON THE VALIDITY OF THE JOSEPHSON HAMILTONIAN}
\label{Hamiltonian}

The original derivation of the S-QPC Hamiltonian in the short junction limit was done in Ref. \cite{Feigelman1999}. It was later shown that for the basis to be phase independent, the correct expression should actually be \cite{Zazunov}:

\begin{equation}
\begin{aligned}
    \hat{H}_\mathrm{} = - \Delta \begin{pmatrix}
        0 & Z \left (r, \frac{\hat{\varphi}}{2} \right ) \\
        Z^* \left (r, \frac{\hat{\varphi}}{2} \right ) & 0 \\
        \end{pmatrix}
\end{aligned}
\label{Zazunov equation}    
\end{equation}

with $Z \left (r,  \frac{\hat{\varphi}}{2} \right ) = e^\mathrm{-i \sqrt{r} \frac{\hat{\varphi}}{2}} \times \left [ \cos \left ( \frac{\hat{\varphi}}{2} \right ) + i \sqrt{r} \sin \left ( \frac{\hat{\varphi}}{2} \right ) \right ]$. Nevertheless, both Hamiltonians have the same eigenvalues and only the basis differs. We have not used the Hamiltonian of Eq. \ref{Zazunov equation} but instead used the one of Eq. \ref{ITLZ H} of the main text, i.e. the one derived in Ref. \cite{Feigelman1999}. We made this choice for two reasons. First in Eq. \ref{Zazunov equation}, the Hamiltonian was derived using the condition of charge neutrality, which is not necessarily met in our graphene junction. Second, this Hamiltonian is not phase periodic, except for some specific values of $r$, which makes its numerical implementation challenging. While the physical meaning of the lack of periodicity is not straightforward, it seems to indicate the loss of charge quantization in the system.

\section{OBTAINING JUNCTION PARAMETERS FROM EXPERIMENTAL DATA}
\label{lookup}

For a given experiment, e.g. a charge dispersion (resp. anharmonicity) measurement, we could extract a pair of experimental data. We obtained a couple $\left ( f_\mathrm{01} \left ( n_\mathrm{g} = 0.25 \right ), \delta f_\mathrm{01} \right )_\mathrm{data}$ (resp. $\left ( f_\mathrm{01}, \alpha \right )_\mathrm{data}$) for each gate voltage $V_\mathrm{g}$. Our diagonalization of Hamiltonian \ref{ITLZ H} of the
main text is numerical, and we have no way to ‘invert the problem’, i.e. to deduce $E_\mathrm{J}$ and $\tau$ values from the eigenergies. This also means that we cannot directly infer these values from the experimental $f_\mathrm{01}$, $f_\mathrm{02}$ (or $\alpha$) and $\delta f_\mathrm{01}$ data. Instead, we build look-up tables, that
make the correspondance between $\left ( E_\mathrm{J}, \tau \right )$ and these quantities. Such tables are obviously of finite size as we compute a finite discrete number of these correspondances, which set the precision on our estimates of $E_\mathrm{J}$ and $\tau$.

\section{ADDITIONAL MEASUREMENTS}
\label{meas}

\subsection{DC electronic transport measurement in a graphene Josephson junction}

\begin{figure}[!htb]
\includegraphics{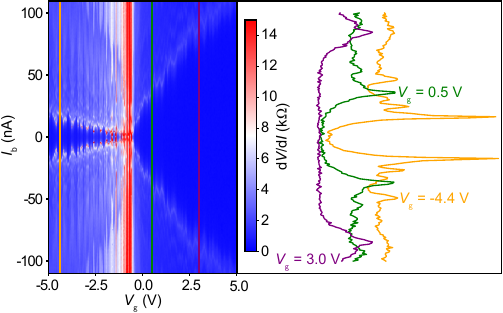}
\caption{\label{fig:DC} Gate-tunable supercurrent in a graphene Josephson junction measured at 25mK. (a) Differential resistance, $dV/dI$, as a function of applied DC current bias, $I_\mathrm{b}$, and topgate voltage, $V_\mathrm{g}$. The critical current is tuned from $\sim 0$ to $\sim 100$ nA for $V_\mathrm{g}$ varying by a few volts. We measure a charge neutrality at negative $V_\mathrm{g}$, $V_\mathrm{CNP} \sim -0.75$ V. (b) Traces at $V_\mathrm{g} = -4.4$ V, $0.5$ V, and $3.0$ V along the colored lines in the left panel.}
\end{figure}

Figure~\ref{fig:DC} shows the differential resistance map and single traces measured in a graphene Josephson junction (of length and width 400 nm, which is similar to the junctions in devices Q$_\mathrm{1}$ and Q$_\mathrm{2}$) with a low-frequency (f = 17.7 Hz) lock-in technique, as a function of current bias $I_\mathrm{b}$ and gate voltage $V_\mathrm{g}$. We observe an asymmetry in the supercurrent about the charge neutrality point of graphene ($V_\mathrm{CNP} \sim - 0.75$ V), due to the $n$-doping by the Ti/Al superconducting electrodes. We notice oscillations in the junction's critical current in the $p$-doped regime of graphene  that are due to the formation of an $npn$ junction in the weak link. This is a typical behaviour that has already been observed in other graphene-based Josephson junctions, a signature of ballistic electronic transport occurring in the junction \cite{Calado2015, Jan2018}.

\subsection{Gate-dependent spectrum of the resonator coupled to a graphene superconducting qubit}

\begin{figure}[!htb]
\includegraphics{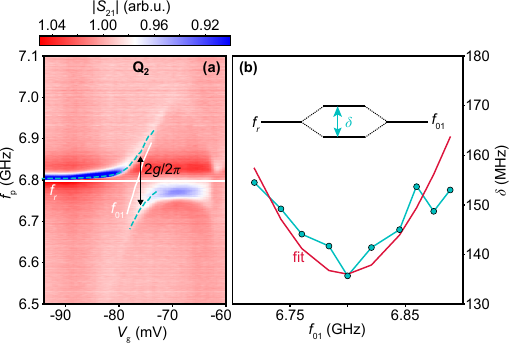}
\caption{\label{fig:1tonVg} Hybridization of the qubit and cavity states in device Q$_\mathrm{2}$. (a) Single-tone transmission measurement of the cavity as a function of probe frequency, $f_\mathrm{p}$, and gate voltage, $V_\mathrm{g}$, applied on the top finger gate. An avoided crossing is apparent close to $V_\mathrm{g} \sim - 75$ mV, which corresponds to vacuum Rabi splitting between the qubit and the cavity. The two solid lines represent the (non-hybridized) bare cavity mode, $f_\mathrm{r} = 6.79$ GHz (white), and the qubit mode, $f_\mathrm{01}$ (white). (b) Splitting, $\delta = f_\mathrm{up} - f_\mathrm{down}$, as a function of $f_\mathrm{01}$. The red line is a fit with Eq. \ref{hybrid eq}. Top inset shows the hybridization of the qubit and cavity states.
}
\end{figure}

The qubit frequency is tunable by the gate voltage, which allows us to control the coupling between the resonator and the qubit. In the vicinity of the resonant limit $f_\mathrm{01} \left ( V_\mathrm{g} \right ) \sim f_\mathrm{r}$, the qubit and resonator hybridize. The $f_\mathrm{up} \left ( V_\mathrm{g} \right )$ and $f_\mathrm{down} \left ( V_\mathrm{g} \right )$ branches show an avoided crossing [Fig. \ref{fig:1tonVg}(a) in device Q$_\mathrm{2}]$. In this regime, the gate-dependent spacing between the branches is given by

\begin{equation}
    \begin{aligned}
        \delta \left ( V_\mathrm{g} \right ) = f_\mathrm{up}\left ( V_\mathrm{g} \right ) - f_\mathrm{down} \left ( V_\mathrm{g} \right ) = \sqrt{\left ( f_\mathrm{01} \left ( V_\mathrm{g} \right ) - f_\mathrm{r} \right )^2 + 4 \left ( \frac{g}{2 \pi} \right )^2}
    \end{aligned}
    \label{hybrid eq}
\end{equation}

with $\frac{g}{2 \pi}$ the coupling strength between the qubit and the resonator. Close to the avoided crossing, we measured the frequencies of the two branches, $f_\mathrm{up}$ and $f_\mathrm{down}$. The qubit frequency is simply given by $f_\mathrm{01} = f_\mathrm{up} + f_\mathrm{down} - f_\mathrm{r}$. Fitting the data using Eq. \ref{hybrid eq} [Fig. \ref{fig:1tonVg}(b)] allows to determine the coupling strength, $\frac{g}{2 \pi} = 76.9 \pm 0.6$ MHz and $68.0 \pm 0.9$ MHz in devices Q$_\mathrm{1}$ and Q$_\mathrm{2}$. We followed Ref. \cite{Koch2007} to design the coupling strength in our devices. Our electrostatic simulations (see Table \ref{tab:capa}) gave us $\frac{g}{2 \pi}$ estimates of 135.1 MHz and 139.0 MHz in devices Q$_\mathrm{1}$ and Q$_\mathrm{2}$, respectively. The factor $\sim 2$ difference may originate from the fact that we use the charge qubit Hamiltonian developped in Ref. \cite{Koch2007} for tunnel junctions to derive $\frac{g}{2 \pi}$, while our experiments revealed that this Hamiltonian is insufficient to properly describe superconducting qubits based on Josephson junctions having high effective transmission.

\subsection{Gate-variation of the anharmonicity in device Q$_\mathrm{1}$}

\begin{figure}[!htb]
\includegraphics{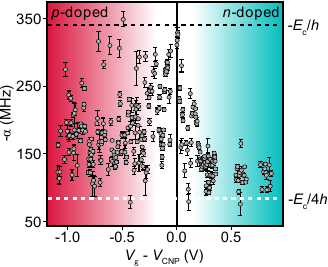}
\caption{\label{fig:anharVg} Anharmonicity in device Q$_\mathrm{1}$ with respect to the gate voltage. The anharmonicity reaches low fractions of $-E_\mathrm{C}/h$ in the $n$-doped regime that is an indication of large effective transmission of the junction. In the hole transport regime, the anharmonicity varies with the gate voltage due to reproducible transmission fluctuations in the weak link.
}
\end{figure}

Figure~\ref{fig:anharVg} shows the anharmonicity extracted in device Q$_\mathrm{1}$ as a function of the gate voltage. For $p$-doped graphene, reproducible anharmonicity fluctuations with the gate voltage are observed, which are reminiscent of what we measured in the DC transport regime [see Fig. \ref{fig:DC}(a)] and in the qubit frequency measurement [see Fig. \ref{fig:energy-tunability}(a)].

\subsection{Charge dispersion measurement in device Q$_\mathrm{2}$}

The charge offset in device Q$_\mathrm{2}$ is controlled by applying a bias voltage $V_\mathrm{j}$ to the superconducting island of the qubit, via the bias tee, on the drive line [see Fig. \ref{fig:device}(a)]. The charge offset is given by $n_\mathrm{g} = \frac{C_\mathrm{d} V_\mathrm{j}}{2e}$, with $C_\mathrm{d} = 168$ aF in this device. The qubit's $\ket{0}$-$\ket{1}$ transition oscillates as a function of $n_\mathrm{g}$. The amplitude of the oscillations gives us the charge dispersion, $\delta f_\mathrm{01} = 613.6 \pm 3.9$ MHz at $V_\mathrm{g}=0.41$ V, and $\delta f_\mathrm{01} = 697.3 \pm 4.2$ MHz at $V_\mathrm{g}=0.44$ V.

\begin{figure}[!htb]
\includegraphics{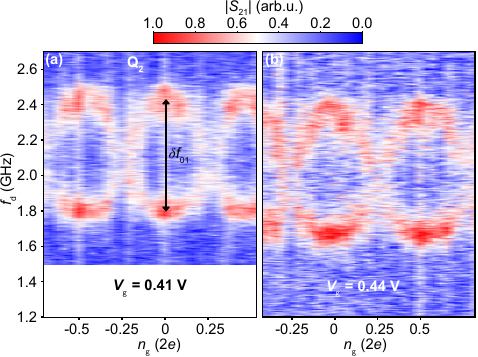}
\caption{\label{fig:deltaf01} Charge dispersion in device Q$_\mathrm{2}$. (a) Two-tone measurement as a function of the charge offset, $n_\mathrm{g}$ (controlled by a voltage applied to the superconducting island of the qubit, $V_\mathrm{j}$), and qubit drive frequency $f_\mathrm{d}$, at $V_\mathrm{g} = 0.41$ V. (b) Measurement performed at $V_\mathrm{g} = 0.44$ V.
}
\end{figure}

\subsection{Estimation of $T_\mathrm{2}^*$ in device Q$_\mathrm{1}$}

\begin{figure}[!htb]
\includegraphics{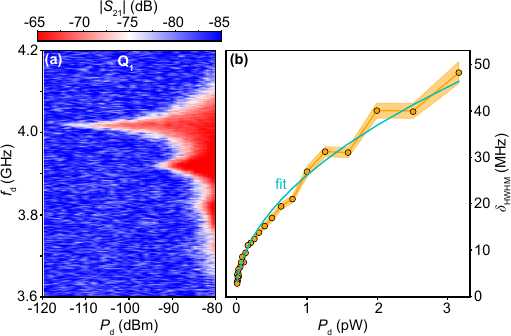}
\caption{\label{fig:2tons} Power-dependence of the energy spectrum of device Q$_\mathrm{1}$, and of the $\ket{0}-\ket{1}$ transition linewidth. (a) Two-tone measurement as a function of qubit drive power, $P_\mathrm{d}$, and qubit drive frequency, $f_\mathrm{d}$. (b) Measured qubit transition linewidth as a function of $P_\mathrm{d}$, and fit to the data.}
\end{figure}

The $\ket{0}$-$\ket{1}$ transition linewidth, $\delta_\mathrm{HWHM}$, is related to the coherence time of the qubit at zero drive power, $T_\mathrm{2}^*$, and the relaxation time of the qubit, $T_\mathrm{1}$, by \cite{Schuster2005, Hertel}

\begin{equation}
    \begin{aligned}
        \left (2 \pi \delta_\mathrm{HWHM} \right )^2 = \left ( \frac{1}{T_\mathrm{2}^*} \right ) + n_\mathrm{s} \omega_\mathrm{vac}^2 \frac{T_\mathrm{1}}{T_\mathrm{2}^*}
    \end{aligned}
    \label{T2 eq power}
\end{equation}

with $n_\mathrm{s} \omega_\mathrm{vac}^2$ a term proportionnal to the drive power on-chip, $\omega_\mathrm{vac} = 2g$ the vacuum Rabi frequency, and $n_\mathrm{s}$ the average number of photons inside the resonator. A two-tone measurement, sweeping the drive frequency at increasing drive power [Fig. \ref{fig:2tons}(a) at $V_\mathrm{g}-V_\mathrm{CNP} = -55$ mV], reveals transitions to the first, second, and third excited state of the qubit's spectrum. The linewidth of the first transition increases with power, as shown in Fig. \ref{fig:2tons}(b). The power dependence of the linewidth is fitted using Eq. \ref{T2 eq power}, and the decoherence time is estimated from the $y$-intercept of the fitted curve. We obtain $T_\mathrm{2}^* = 62.0 \pm 14.5$ ns, a relatively short value that could be explained by charge noise (for this particular working point close to the charge neutrality point of graphene, the $E_\mathrm{J}/E_\mathrm{C}$ ratio, the charge dispersion is expected to be on the order of 3.7 MHz), or Josephson energy noise induced by the gate line.

\subsection{Relaxation time $T_\mathrm{1}$ measurement in device Q$_\mathrm{1}$}

\begin{figure}[!htb]
\includegraphics{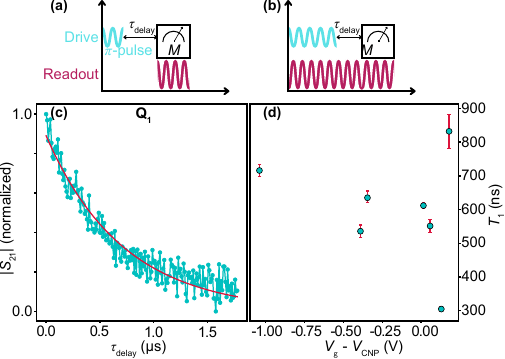}
\caption{\label{fig:T1} Relaxation time measurement in device Q$_\mathrm{1}$. (a) Standard $T_\mathrm{1}$-measurement protocol. A $\pi$-pulse is sent to the qubit to excite state $\ket{1}$ and a readout tone is sent after $\tau_\mathrm{delay}$ to measure the exponential decay of the qubit population. (b) Overlapping pulses sequence. (c) $T_\mathrm{1}$-measurement at $V_\mathrm{g} - V_\mathrm{CNP} = -1.04$ V. $T_\mathrm{1}$ is extracted by a fit (solid line) with a $A e^{-\frac{\tau_\mathrm{delay}}{T_\mathrm{1}}}$ function. (d) Extracted $T_\mathrm{1}$ as a function of $V_\mathrm{g} - V_\mathrm{CNP}$.}
\end{figure}

We did not manage to measure the decay time of the qubit using the standard scheme illustrated in Fig.~\ref{fig:T1}(a). We  thus followed Ref. \cite{Hertel} and used overlapping $\sim 1$ $\mu$s-long readout and drive pulses. The drive and readout pulses are sent at the same time, and the demodulation sequence (white box labeled $M$) starts after a delay $\tau_\mathrm{delay}$, as shown in Fig.~\ref{fig:T1}(b) . The result from such a measurement is presented in Fig.~\ref{fig:T1}(c) at $V_\mathrm{g} - V_\mathrm{CNP} = -1.04$ V. We extracted a $T_\mathrm{1}$-value of 716 $\pm$ 17.4 ns, which is one order of magnitude longer than what was reported for graphene gatemons in Ref. \cite{Jan2018}. Figure~\ref{fig:T1}(d) shows the gate-dependence of $T_\mathrm{1}$, where no clear pattern can be observed. The variation of $T_\mathrm{1}$ with $V_\mathrm{g}$ might be an indication that $V_\mathrm{g}$-dependent dissipation occurs in the junction or that the interactions with the environment depend on the qubit frequency. Leakage through the gate could be relevant to this respect. Further measurements, beyond the scope of the present work, are needed to understand the main factors limiting the decay times in our circuits.

\bibliography{Biblio_paper}

\end{document}